\documentstyle[12pt]{article}
\begin{document}
\title{Superconvergent Perturbation Method in Quantum Mechanics}
\author{Wolfgang Scherer \\ Institut f\"ur Theoretische Physik A \\
                           TU Clausthal \\ Leibnizstr. 10 \\
                           D--38678 Clausthal--Zellerfeld \\ Germany}
\date{}
\maketitle
\begin{abstract}
An analogue of Kolmogorov's superconvergent perturbation theory in
classical mechanics is constructed for self adjoint operators. It
is different from the usual Rayleigh--Schr\"odinger perturbation
theory and
yields expansions for eigenvalues and eigenvectors in terms
of functions of the perturbation parameter.
\end{abstract}
\vskip5mm
{\it 1. Introduction.}
About a century ago Lindstedt~\cite{Lindst}, Poincar\'e~\cite{Poinca},
and von Zeipel~\cite{Zeipel}
developed a perturbation theory for Hamiltonian systems in classical
mechanics (CM). The method was mainly used in celestial mechanics but
in most cases failed to converge due to the appearance of
small divisors. It was only in the fifties that
Kolmogorov~\cite{Kolmog} proposed a new quadratically convergent
perturbation method for Hamiltonian systems in CM and this
(nowadays called superconvergent) method was essential in Arnold's
proof of the KAM theorem~\cite{Arnold}.

In this letter an exact analogue of Kolmogorov's superconvergent
method is constructed for self adjoint operators. So far
``superconvergent"is only a name for this new method and a detailed
functional analytic investigation will have to determine how much
the new perturbation theory constructed here is an improvement
on existing schemes. There are, however, indications that this will
be so: The first comes from the fact that we have also constructed
an analogue of the Poincar\'e--von Zeipel perturbation theory
for self adjoint operators and have shown that this analogue is
identical to the usual Rayleigh--Schr\"odinger perturbation
theory~\cite{QA1}. Since in CM the superconvergent method is a vast
improvement on the Poincar\'e--von Zeipel method we may expect
the same in Quantum Mechanics (QM). The second
comes from initial numerical studies.

{\it 2. General Algorithm of the Superconvergent Method.}
Let $H^0_0$ be the unperturbed Hamiltonian (CM: function on phase
space, QM: operator) and let
\begin{equation}
\label{e1}
H^0(\epsilon):=\sum_{p=0}^\infty \frac{\epsilon^p}{p!} H^0_p
\end{equation}
be the perturbed Hamiltonian where the $H^0_p$ do not depend on
the perturbation parameter $\epsilon$. Moreover, let
\begin{equation}
\label{e2}
W^n(\epsilon):=\sum_{p=0}^\infty \frac{\epsilon^p}{p!} W^n_{p+1}
\end{equation}
and let $-W(\epsilon)$
be generators (CM: functions on phase space, QM: operators) of
(CM: canonical, QM: unitary) flows $(\Phi^n(\epsilon))^{-1}$
 with ``time" $\epsilon$.
This means that the transformations $\Phi^n(\epsilon)$
satisfy the initial value problem
\begin{equation}
\label{e3}
\frac{d}{d\epsilon}\Phi^n(\epsilon)^*  =
ad\; W^n(\epsilon) \circ \Phi^n(\epsilon)^*,
\qquad  \Phi^n(0)  =  id
\end{equation}
where in CM the $\Phi^n(\epsilon)^*$ act on any phase space function $A$,
via
\begin{equation}
\Phi^n(\epsilon)^*(A):= A\circ \Phi^n(\epsilon)
\end{equation}
and where in CM
\begin{equation}
\label{e4}
ad\, W^n(\epsilon)(A):=\{ W^n(\epsilon), A\},
\end{equation}
and $\{\cdot ,\cdot\}$ denotes the Poisson bracket.
In QM self adjoint operators $-W^n(\epsilon)$ generate one parameter
groups of unitary transformations $(\Phi^n(\epsilon))^{-1}$
such that the $\Phi^n(\epsilon)$  solve
the same initial value problem as given in (\ref{e3})
but where the $\Phi^n(\epsilon)^*$
act on operators $A$ via
\begin{equation}
\label{e5}
\Phi^n(\epsilon)^*(A):= \Phi^n(\epsilon)^{-1}\; A\; \Phi^n(\epsilon)
\end{equation}
and where now
\begin{equation}
\label{e6}
ad\, W^n(\epsilon)(A):=\frac{i}{\hbar}[ W^n(\epsilon), A ]
\end{equation}
and $[\cdot ,\cdot ]$ denotes the commutator. With the understanding
that the appropriate definition (\ref{e4}) or (\ref{e6}) (depending
on whether one is dealing with CM or QM) is chosen we
proceed to present Kolmogorov's superconvergent method.
In the followng it will be useful to have an expansion for
$\Phi^n(\epsilon)^*$ in terms of operators $T^n_p$ independent
of $\epsilon$. Writing
\begin{equation}
\label{e8}
\Phi^n(\epsilon)^* =
\sum_{p=0}^\infty \frac{\epsilon^p}{p!} T^n_p
\end{equation}
one finds that the $T^n_p$ are recursively determined by $T^n_0=id$
and
\begin{equation}
\label{e9}
T^n_{p+1}=\sum_{l=0}^p {p \choose l}\,ad\, W^n_{l+1} \circ T^n_{p-l}.
\end{equation}
In CM the $T^n_p$ are differential operators such that $T^n_p(A)$
means application of $T^n_p$ to the phase space function $A$ whereas
in QM they are operators acting on operators.

Kolmogorov's method consists of finding the generators $W^n(\epsilon)$
in such a way that
\begin{equation}
\label{e10}
K^{n-1}(\epsilon) := \Phi^{n-1}(\epsilon)^*\circ\Phi^{n-2}(\epsilon)^*
\circ\cdots\circ\Phi^1(\epsilon)^* (H^0(\epsilon))
\end{equation}
has an expansion in $\epsilon$
\begin{equation}
\label{e11}
K^{n-1}(\epsilon):=\sum_{p=0}^\infty\frac{\epsilon^p}{p!} K^{n-1}_p
\end{equation}
with ``integrable" terms up to order $2^{n-1}-1$
(whereas in CM the term integrable has a well defined meaning this fails
to be the case in QM; however, we only use the notion of integrability
to motivate the
superconvergent method, all its equations are well defined in CM without
any need to define integrability and will be well defined operator
equations in QM even though the term integrability is not).
Then a new perturbed Hamiltonian $H^{n-1}(\epsilon):=K^{n-1}(\epsilon)$
is defined such that its unperturbed part $H^{n-1}_0$ consists of the
integrable part of $K^{n-1}$ and its perturbation
is of the order $2^{n-1}$.
 We can summarize the iterative procedure as follows (for a more detailed
discussion we refer the reader to~\cite{LicLie} with the warning that
our notation differs from theirs): \newline
\[
\begin{tabular}{l|l|l}
result of $\Phi^{n-1}$:
& choice of $\Phi^n$:
& result of $\Phi^n$:  \\
$H^{n-1}$   &  $W^n$  & $H^n$  \\ \hline
$H^{n-1}_0$,  &
$W^n_p=0$, \hfill $1\le p<2^{n-1}$ &
$H^n_0=H^{n-1}_0$  \\
(int. up to $O(2^{n-1}-1)$) &
$ad\,W^n_p(H^{n-1}_0)=$ &
\hspace*{2em}$+\sum_{p=2^{n-1}}^{2^n-1}\frac{\epsilon^p}{p!}
\overline{H^{n-1}_p}$,\\
  &
\hspace{2em}$\overline{H^{n-1}_p}-H^{n-1}_p$, \hfill
$2^{n-1}\le p < 2^n$ &
(int. up to $O(2^n-1)$) \\
$H^{n-1}_p=0$, \hfill $1\le p<2^{n-1}$ &
with $ad\, H^{n-1}_0(\overline{H^{n-1}_p})=0$ &
 \\
(no perturbation) &
$W^n_p=0$, \hfill $2^n\le p$ &
 \\  \cline{2-2}
  &
leads to &
$H^n_p=0$, \hfill $1\le p < 2^n$ \\  \cline{2-2}
$H^{n-1}_p=K^{n-1}_p$, \hfill $2^{n-1}\le p$ &
$K^n_0=H^{n-1}_0$ &
(no perturbation) \\
 &
$K^n_p=0$, \hfill $1\le p <2^{n-1}$ &
  \\
 &
$K^n_p=\overline{H^{n-1}_p}$, \hfill $2^{n-1}\le p <2^n$ &
$H^n_p=K^n_p$, \hfill $2^n\le p$ \\
 &
$K^n_p=H^{n-1}_p$ &
 \\
 &
\hspace*{2em}$+\sum_{j=1}^{p-1}{p-1 \choose j-1}\left( ad\,
W^n_j(K^n_{p-j}) \right.$
 &
 \\
 &
\hspace*{3em}$\left.+ T^n_{p-j}(H^{n-1}_j)\right)$,  \hfill
$2^n\le p$ &
 \\
\end{tabular}
\]
The meaning of $\overline{\cdot}$ is to be understood as follows.
Assume $H^{n-1}$ is of the form given in the left column and choose
$W^n_p=0$ for $1\le p <2^{n-1}$. Then one finds first
$K^n_0  =  H^{n-1}_0,\;
K^n_p  =  0$, for $1 \le p < 2^{n-1}$ and
\begin{equation}
K^n_p  =  ad\, W^n_p (H^{n-1}_0) + H^{n-1}_p\, ,
\qquad 2^{n-1}\le p < 2^n. \label{kneq}
\end{equation}
Since $H^{n-1}_0$ is already integrable up to $O(2^{n-1}-1)$ and $\Phi^n$
should improve this, one would like to have that the $K^n_p$ for
$2^{n-1}\le p < 2^n$ are integrable.
Consequently the crucial point in this procedure
becomes the construction of the $W^n_p$ and some $\overline{H^{n-1}_p}$
for $2^{n-1}\le p < 2^n$ such that
\begin{equation}
\label{e26}
ad\, W^n_p(H^{n-1}_0)=\overline{H^{n-1}_p}-H^{n-1}_p
\end{equation}
where $\overline{H^{n-1}_p}$ is such that
\begin{equation}
\label{e27}
ad\, H^{n-1}_p (\overline{H^{n-1}_p}) =0.
\end{equation}
Then it follows from (\ref{kneq}) that
$K^n_p=\overline{H^{n-1}_p}$ and from~(\ref{e27}) that they commute
with the unperturbed Hamiltonian $H^{n-1}_0$ of the previous step.
The table summarizing the method omits these
intermediate steps and shows only the results once (\ref{e26}) and
(\ref{e27}) have been solved.

In CM (\ref{e26}) and (\ref{e27}) can be satisfied with the help
of the averaging method and $\overline{H^{n-1}_p}$ turns out to be
the average of $H^{n-1}_p$ over the angle variables of the tori
which are assumed to be compact.
It should be noted, however, that the
$\overline{H^{k-1}_p}$ in general depend on $\epsilon$ such that
(\ref{e25}) is actually an expansion in terms of functions of
the perturbation parameter.
Since $H^{n-1}_0$ is assumed to be
integrable (\ref{e27}) implies that $\overline{H^{n-1}_p}$
and hence $H^{n}_0$ are integrable as well.

Hence, after the $n$th transformation one has a Hamiltonian
$H^n(\epsilon)$ where
\begin{equation}
\label{e25}
H^n_0=\sum_{k=0}^n\left(\sum_{p=2^{k-1}}^{2^k-1}\frac{\epsilon^p}{p!}
\overline{H^{k-1}_p}\right)
\end{equation}
is integrable and the perturbations $H^n_p$ are
of order $2^n$ or higher.

{\it 3. Superconvergent Method for Self Adjoint Operators.}
With our previous notation equations (\ref{e26}) and (\ref{e27}) become
operator equations in QM which have to be solved for each step.
We can construct operators
$W^n_p, \overline{H^{n-1}_p}$ satisfying (\ref{e26}) and (\ref{e27})
with the help of the following quantum analogue of the classical
averaging procedure~\cite{QA1} which is a modified
version of an idea used by Weinstein~\cite{Weinst} in the
context of pseudodifferential operators:
Let $A,B$ be self adjoint operators such that
\begin{eqnarray}
B^{(A)}(t) & := &
\exp(-\frac{i}{\hbar}tA)\, B \, \exp(\frac{i}{\hbar}tA) \\
\overline{B}^{(A)} & := & \lim_{T\to\infty}\frac{1}{T}
\int_0^T\, B^{(A)}(t) dt \label{condB} \\
S^{(A)}(B) & := & \lim_{T\to\infty}\frac{1}{T}
\int_0^T dt\,\int_0^t ds\, \left( B^{(A)}(s) -\overline{B}^{(A)}\right)
\label{condS}
\end{eqnarray}
exist and
\begin{equation}
\label{cond}
\lim_{T\to\infty}\frac{B^{(A)}(T)-B}{T} =0
\end{equation}
then the following holds
\begin{eqnarray}
ad\; \overline{B}^{(A)}\;(A) & = &
\frac{i}{\hbar}[\overline{B}^{(A)}, A ]      =  0
\qquad \mbox{and}     \label{e29} \\
ad\; S^{(A)}\;(A) & = & \frac{i}{\hbar}[S^{(A)}(B),A]
=  \overline{B}^{(A)}-B.  \label{e28}
\end{eqnarray}
{\it Proof:}
\begin{eqnarray*}
ad\; \overline{B}^{(A)}\;(A) & = & \lim_{T\to\infty}\frac{1}{T}
\int_0^T \frac{i}{\hbar}[B^{(A)}(t),A] dt =
\lim_{T\to\infty}\frac{1}{T}
\int_0^T \frac{d}{dt}B^{(A)}(t) dt \\
\mbox{} & = &  \lim_{T\to\infty}\frac{B^{(A)}(T)-B}{T} = 0
\end{eqnarray*}
by assumption and thus
\begin{eqnarray*}
ad\; S^{(A)}(B)\; (A) & = &
\lim_{T\to\infty}\frac{1}{T}
\int_0^T dt\int_0^t ds\, \frac{i}{\hbar}[B^{(A)}(s),A]  =
\lim_{T\to\infty}\frac{1}{T}
\int_0^T dt \int_0^t ds\, \frac{d}{ds}B^{(A)}(s) \\
\mbox{} & = & \lim_{T\to\infty}\frac{1}{T}
\int_0^T dt \, \left( B^{(A)}(s) - B\right) = \overline{B}^{(A)} - B
\end{eqnarray*}
which completes the proof.
Setting now
$W^n_p= S^{(H^{n-1}_0)}(H^{n-1}_p)$
solves (\ref{e26}) and (\ref{e27}) in QM with
$\overline{H^{n-1}_p}=\overline{H^{n-1}_p}^{(H^{n-1}_0)} =:
\overline{H_p}^{(n-1)}$
where the last equation introduces a simplified notation.

{\it 4. Interpretation of the Algorithm for Self Adjoint Operators.}
The results of section~3 show that we can execute
Kolmogorov's superconvergent
perturbation algorithm in QM as well. But what has been gained?
By construction $H^n(\epsilon)$ and the original perturbed
Hamiltonian $H^0(\epsilon)$ are unitarily equivalent:
\begin{equation}
\label{unit}
H^n(\epsilon)=\Phi^n(\epsilon)^{-1}\cdots
\Phi^1(\epsilon)^{-1}\, H^0(\epsilon) \,
\Phi^1(\epsilon)\cdots
\Phi^n(\epsilon)
\end{equation}
and thus have
identical spectra. Since, after the first transformation
$[H^0_0,\overline{H_1}^{(0)}]=0$, we can diagonalize $H^0_0$
and $\overline{H_1}^{(0)}$ simultanously which permits us to
diagonalize $H^1_0:=H^0_0+\epsilon\overline{H_1}^{(0)}$.
After the second transformation we have
\begin{equation}
[H^1_0,\overline{H_2}^{(1)}]=0=[H^1_0,\overline{H_3}^{(1)}]
\end{equation}
and thus we can diagonalize
\begin{equation}
H^2_0=H^0_0+\epsilon \overline{H_1}^{(0)}+\frac{\epsilon^2}{2!}
\overline{H_2}^{(1)} + \frac{\epsilon^3}{3!}\overline{H_3}^{(1)}
\end{equation}
etc. In this way we arrive after n transformations at a diagonal
$H^n_0(\epsilon)$ whose eigenvalues $E^n_j(\epsilon)$ approximate
those of $H^n(\epsilon)$ and thus of $H^0(\epsilon)$.
The eigenvectors $|j\rangle^n$ of $H^n_0$ are known by construction
and they approximate those of $H^n$. Because of (\ref{unit}) it
follows that
$\Phi^1(\epsilon)\cdots
\Phi^n(\epsilon)|j\rangle^n$ are the appropriately
approximated eigenvectors of the original perturbed $H^0(\epsilon)$.
In formulae: If
$H^0(\epsilon)\,|j\rangle (\epsilon) = E_j(\epsilon)\, |j\rangle (\epsilon)$
then
\begin{eqnarray}
E_j(\epsilon) & = & E^n_j(\epsilon) + O(2^n)  \\
|j\rangle (\epsilon) & = & \Phi^1(\epsilon)\cdots
\Phi^n(\epsilon) |j\rangle^n + O(2^n).
\end{eqnarray}
Since the
$\overline{H_p}^{(n)}$ depend on $\epsilon$ the $E^n_j(\epsilon)$
are no longer a pure power series expansion but should be viewed
as an expansion in terms of functions of $\epsilon$.
These functions themselves have in general infinite series
expansions in $\epsilon$ so that already low orders $E^n_j(\epsilon)$
contain contributions of all orders in $\epsilon$. The example
treated in the next section will make this point clearer.

{\it 5. An Example: Discrete and Nondegenerate Spectrum.}
It remains to show that the method constructed her is truly distinct
from the usual Rayleigh--Schr\"odinger perturbation theory.
For this purpose we sketch
the results for the case of a Hamiltonian
$H^0_0=\sum_{j}\,|j\rangle\,E^{0}_j\;\langle j|$
with purely
discrete and nondegenerate spectrum and a perturbation
only linear in $\epsilon$, i.~e.~
$H^0_1=\sum_{j,k}\, |j\rangle \; V_{jk} \; \langle k|$
and $H^0_p=0$ for $p\ge 2$.
In this case the eigenvectors $|j\rangle^1$ of $H^1_0=H^0_0+
\epsilon \overline{H_1}^{(0)}$ and
$|j\rangle$ of $H^0_0$ coincide because $H^0_0$ and
$\overline{H_1}^{(0)}$ commute and $H^0_0$ has nondegenerate
spectrum. Continuing this argument inductively it is easy to see that
$H^0_0$ and all $H^n_0$ have the same eigenvectors $|j\rangle$.
By straightforward calculation one finds
\begin{eqnarray}
W^n_p & = & \frac{\hbar}{i} \sum_{j\neq k} |j\rangle
\frac{\langle j| H^{n-1}_p |k\rangle}{E^{n-1}_j-E^{n-1}_k}\langle k| \\
\overline{H_p}^{(n-1)} & = &
\sum_j |j\rangle \langle j|H^{n-1}_p |j\rangle
\langle j| \\
E^n_j & = & E^{0}_j + \sum_{k=1}^n\left(
\sum_{l=2^{k-1}}^{2^k-1}\frac{\epsilon^l}{l!} E^{(l)}_j \right)
\end{eqnarray}
where $E^{(l)}_j:=\langle j|H^{k-1}_l|j\rangle,\; 2^{k-1}\le l\le2^k-1$
and the $E^n_j$ are the eigenvalues of $H^n_0$.
{}From the algorithm presented in section 2
one obtains
\begin{eqnarray}
H^1_2 & = & \frac{i}{\hbar}[W^1_1, \overline{H_1}^{(0)} + H^0_1 ]  \\
H^1_3 & = & \left(\frac{i}{\hbar}\right)^2
[W^1_1, [W^1_1, \overline{H_1}^{(0)} + 2 H^0_1 ] ]  \\
H^2_4 & = & \left(\frac{i}{\hbar}\right)^3
[ W^1_1, [ W^1_1, [ W^1_1, \overline{H_1}^{(0)} + 3 H^0_1 ]]]
+ 3 \frac{i}{\hbar}[W^2_2, \overline{H_2}^{(1)} + H^1_2 ].
\end{eqnarray}
The appearance of the denominators $E^{n-1}_j-E^{n-1}_k$ in $W^n_p$
leads to a perturbation theory substantially different from the usual
Rayleigh--Schr\"odinger theory. However, this difference only shows
up in the fourth and higher order terms
since in $H^1_2$ and $H^1_3$ only commutators with $W^1_1$ appear which
contains only the usual Rayleigh--Schr\"odinger
denominator $E^0_j-E^0_k$. Indeed, one finds
\begin{eqnarray}
E^{(1)}_j & = & V_{jj}  \\
\frac{1}{2!}E^{(2)}_j&=&\sum_{j\neq k} \frac{|V_{jk}|^2}{E^0_j-E^0_k} \\
\frac{1}{3!}E^{(3)}_j&=&\sum_{m\neq j\neq k}
\frac{V_{jk}V_{km}V_{mj}}{(E^0_j-E^0_k)(E^0_j-E^0_m)}
-\sum_{j\neq k}\frac{|V_{jk}|^2V_{jj}}{(E^0_j-E^0_k)^2}
\end{eqnarray}
showing that up to $O(3)$ the new method coincides with the usual
Rayleigh--Schr\"odinger perturbation theory. But $H^2_4$
contains $W^2_2$ with denominators of the form
$E^1_j-E^1_k=E^0_j-E^0_k + \epsilon (V_{jj}-V_{kk})$
which are functions of $\epsilon$ and which appear as denominators in
\begin{eqnarray*}
E^{(4)}_j  =
&24& \sum_{j\neq l}
\frac{|V_{lj}|^2(V_{ll}-V_{jj})^2}{(E^0_j-E^0_l)^2(E^1_j-E^1_l)} \\
&+& 6\sum_{j\neq l\neq k\neq j\neq m\neq k}
\frac{V_{jl}V_{lk}V_{km}V_{mj}}{E^1_j-E^1_k}
\left( \frac{1}{E^0_j-E^0_l} - \frac{1}{E^0_l-E^0_k} \right)
\left( \frac{1}{E^0_k-E^0_m} - \frac{1}{E^0_m-E^0_j} \right)  \\
&+& 12\sum_{j\neq l\neq k\neq j} V_{jl}V_{lk}V_{kj}\left\{
\frac{V_{ll}-V_{kk}}{(E^0_j-E^0_l)(E^0_l-E^0_k)(E^0_k-E^0_j)} \right. \\
&\mbox{}& \hspace*{9em} +
\frac{V_{ll}-V_{jj}}{(E^0_j-E^0_l)(E^1_j-E^1_l)}
\left(\frac{1}{E^0_l-E^0_k}-\frac{1}{E^0_k-E^0_j}\right) \\
&\mbox{}& \hspace*{10em} +  \left.
\frac{V_{kk}-V_{jj}}{(E^0_j-E^0_k)(E^1_j-E^1_k)}
\left(\frac{1}{E^0_j-E^0_l}-\frac{1}{E^0_l-E^0_k}\right) \right\} \\
&+& \sum_{k\neq l\neq j\neq m\neq k} V_{jl}V_{lk}V_{km}V_{mj}
\left\{
\frac{9}{(E^0_m-E^0_j)(E^0_j-E^0_l)}
\left( \frac{1}{E^0_k-E^0_m} - \frac{1}{E^0_l-E^0_k} \right) \right. \\
\mbox{} &\mbox{}& \hspace*{9em}  \left.
+\frac{3}{(E^0_l-E^0_k)(E^0_k-E^0_m)}
\left(\frac{1}{E^0_j-E^0_l} - \frac{1}{E^0_m-E^0_j} \right)
\right\}
\end{eqnarray*}
As an example we quote the result for the ground state correction
up to $E^{(4)}_0$ for the quartic perturbation $H^0_1=x^4$ of the
harmonic oscillator $H^0_0 =-\frac{d^2}{dx^2} + x^2$ (with $\hbar=1$)
\begin{eqnarray*}
E^0_0(\epsilon)_{SU}
& = & 1 + \frac{3}{4}\epsilon - \frac{21}{16}\epsilon^2
+ \frac{333}{64}\epsilon^3  \\
\mbox{} & - &
\frac{3(1317760 + 12935472 \epsilon + 36433368 \epsilon^2 +
25183305 \epsilon^3)}{2048(4+9 \epsilon)(4+15 \epsilon)(4+21 \epsilon)}
\epsilon^4 + O(\epsilon^5)
\end{eqnarray*}
which approximates the numerically computed eigenvalues~\cite{GaPaII}
much better
than the standard fourth order Rayleigh--Schr\"odinger correction
\begin{equation}
E^0_0(\epsilon)_{RS}
= 1 + \frac{3}{4}\epsilon - \frac{21}{16}\epsilon^2
+\frac{333}{64}\epsilon^3 -\frac{30885}{1024}\epsilon^4 +O(\epsilon^5).
\end{equation}
This example also illustrates the statements made earlier that the
lower order corrections $E^{(n)}_j$ already contain infinite power
series of $\epsilon$ which may be the reason for improved
convergence.

{\it 6. Conclusion.}
It should be emphasized that the method presented here is very
general and in principle applicable to any self adjoint operator $A$
and perturbation $B$ provided they (and the higher order operators)
satisfy the conditions needed to construct $\overline{B}^{(A)}$ and
$S^{(A)}(B)$ in~(\ref{condB}) and~(\ref{condS}).
It is also designed to include analytic perturbations which are not
necessarily only linear in the perturbation parameter.
Moreover, it gives the corrections as integrals of the form
$\overline{B}^{(A)}$ which, if desired, can be exhibited as sums
over intermediate states as was done in section~5, but
which may be evaluated directly circumventing the calculation
of these sums which may sometimes be impossible. This was
already shown to be an advantage of the formulation of the
usual Rayleigh--Schr\"odinger theory as an analogue of the
classical Poincar\'e--von Zeipel perturbation method in~\cite{QA1}.

The method will be presented in more detail
in a forthcoming longer
paper~\cite{QA2}.

\end{document}